\documentclass[11pt]{revtex4-2}

\usepackage{graphicx}% Include figure files
\usepackage{dcolumn}% Align table columns on decimal point
\usepackage{bm}% bold math
\usepackage[table]{xcolor}
\usepackage{graphicx}
\usepackage{adjustbox}
\usepackage{placeins}
\usepackage[T1]{fontenc}
\usepackage{lipsum}
\usepackage{csquotes}
\usepackage{colortbl}
\usepackage{hyperref}
\usepackage{bm}
\usepackage{multirow}
\usepackage{booktabs}
\usepackage{epsfig}
\usepackage{amsmath,amssymb}

\begin{document}

\title{Transferable Machine Learning Potential X-MACE for Excited States using Integrated DeepSets}

\author{Rhyan Barrett}
\affiliation{Leipzig University, Wilhelm Ostwald Institute for Physical
and Theoretical Chemistry, Linnéstraße 2, 04103 Leipzig, Germany}
\author{Christoph Ortner}
\affiliation{Mathematics Department, University of British Columbia, 1984 Mathematics Rd, Vancouver, BC V6T 1Z2, Canada}
\author{Julia Westermayr}
\email{julia.westermayr@uni-leipzig.de}
\affiliation{Leipzig University, Wilhelm Ostwald Institute for Physical
and Theoretical Chemistry, Linnéstraße 2, 04103 Leipzig, Germany}
\affiliation{Center for Scalable Data Analytics and Artificial Intelligence (ScaDS.AI), Dresden/Leipzig, Germany}

\date{\today}% It is always \today, today,

\begin{abstract}
Conical intersections serve as critical gateways in photochemical reactions, enabling rapid nonradiative transitions between potential energy surfaces that underpin fundamental processes such as photosynthesis or vision. Their calculation with quantum chemistry is, however, extremely computationally intensive and their modeling with machine learning poses a significant challenge due to their inherently non-smooth and complex nature. To address this challenge, we introduce a deep learning architecture designed to precisely model excited states and improve their accuracy around these critical, non-smooth regions. Our model integrates Deep Sets into the Message Passing Atomic Cluster Expansion (MACE) framework resulting in a smooth representation of the non-smooth excited-state potential energy surfaces. We validate our method using numerous molecules, showcasing a significant improvement in accurately modeling the energy landscape around conical intersections compared to conventional excited-state models. Additionally, we apply ground-state foundational machine learning models as a basis for excited states. By doing so, we showcase that the developed model is capable of transferring not only from the ground state to excited states, but also within chemical space to molecular systems beyond those included in the training dataset. This advancement not only enhances the fidelity of excited-state modeling, but also lays the foundations for the investigation of more complex molecular systems.

\end{abstract}

\maketitle
%%%FOOTNOTES%%%

\section{Introduction}

Photodynamics simulations play a pivotal role in advancing our understanding of essential photophysical and photochemical processes, including but not limited to photosynthesis\cite{hustings2022charge,xing2023exploring}, photo-damage \cite{westermayr_2022_deep,ashfold2006role,yamijala2022photo}, or optoelectronics \cite{wright2021ultrafast,huff2021excited,westermayr2023high, romero2017quantum}. By investigating how molecules absorb, emit, and interact with light, these studies can unveil fundamental mechanisms that underpin key biological functions or disease pathways \cite{jaiswal2020first,dedonder2013excited, kleinermanns2013excited} and can help a more informed design of functional materials. For instance, detailed insights into the energy transfer mechanisms involved in photosynthesis can inform the design of artificial photosynthetic systems \cite{imahori2010photoinduced, manbeck2017tetra}, thereby contributing to innovations in renewable energy technologies. Similarly, a deeper understanding of the interactions of light and matter at the cellular and molecular levels has the potential to improve photodynamics cancer therapies \cite{cwiklik2020bodipy, alvarez2024current} or enhance treatments for vision-related disorders. \cite{valverde2022ultrafast} %These theoretical frameworks not only deepen our comprehension of natural phenomena but also pave the way for breakthroughs across multiple fields, including medicine, energy, and materials science.

In simulating excited state processes, however, the fundamental assumption that underlies many ground-state studies, which is the Born–Oppenheimer approximation, breaks down and complicates excited-state simulations. In regions, where electronic transitions play a crucial role, the separation of nuclear and electronic degrees of freedom is no longer valid. Instead, the electrons and nuclei become tightly coupled, necessitating methodologies that can handle the simultaneous evolution of both.\cite{nelson2022modeling} Mixed quantum-classical molecular dynamics simulations go beyond the Born-Oppenheimer approximation, enabling the study of excited-state dynamics. These simulations treat nuclei classically and propagate them on multiple adiabatic potential energy surfaces, allowing for electronic transitions between them and maintaining computational feasibility while capturing nonadiabatic effects. .\cite{baer2002introduction} An example of such a technique is surface hopping that describes transitions between electronic states as so-called hops. \cite{mai2020surface, mai2015general, sholl1998generalized,tully1991nonadiabatic} %enable the propagation of nuclear dynamics on multiple adiabatic potential energy surfaces while allowing for electronic transitions between them. This approach accounts for the coupling between electronic and nuclear motions, enabling the simulation of processes like radiationless decay and photoisomerization, which are critical in photochemical reactions.

Still,  such techniques remain computationally expensive as they rely on quantum chemical calculations of multiple potential energy surfaces including corresponding gradients at every time step during a dynamics simulation.\cite{westermayr2019machine} The computational cost of such simulations scales dramatically with system size and simulation time \cite{mukherjee2022simulations,Westermayr2021CR}. It is not uncommon for these simulations to require years of computational time to model even small molecules consisting of a few dozen atoms over timescales of hundreds of picoseconds.\cite{Westermayr2021CR,Li2022ACR,westermayr2022deep,zhang} This computational bottleneck severely limits the practical applicability of these methods to larger, more complex systems and longer simulation times, which are often of significant interest in both fundamental research and practical applications.

Machine learning offers a transformative solution to this computational challenge. By constructing an initial dataset of molecular structures and their associated properties, machine learning models can be trained to predict the behavior of larger, more complex systems at significantly faster rates while retaining ab initio accuracy \cite{westermayr2020machine, westermayr2022deep,zhang2024mlatom,li2021automatic}. This approach has demonstrated impressive success in recent years, particularly with the development of neural networks designed for predicting molecular properties for molecuar dynamics. \cite{schutt2018schnet, deng2023chgnet, smith2017ani} Equivariant neural networks, which respect the symmetries inherent in molecular systems such as rotational and translational invariance, have achieved high accuracy and generalization capabilities for ground-state properties.\cite{schutt2021equivariant, musaelian2023learning, batzner20223}

Building on these advances, several neural network architectures, including SPaiNN \cite{mausenberger2024spainn}, SchNarc \cite{westermayr2020combining}, PyrAIMD$^2$\cite{li2021automatic} or MLAtom \cite{zhang2024mlatom, dral2024mlatom, dral2022mlatom}, have been proposed for excited-state prediction, allowing for excited-state simulations in the nanosecond regime.\cite{westermayr2019machine,li2021automatic} While these methods have shown great promise, they face significant challenges in accurately modeling conical intersections in the molecular potential energy landscape where two or more Born–Oppenheimer potential energy surfaces intersect. The non-smooth nature of conical intersections makes them difficult for neural networks to approximate without employing a prohibitively large number of parameters. This is because neural networks are inherently smooth function approximators and struggle to capture sharp, discontinuous features in the data.\cite{Westermayr2021CR, gutleb2024parameterizing}

In this paper, we present a deep learning architecture designed to overcome the challenges associated with excited-state dynamics, particularly in the vicinity of conical intersections. Our model leverages the concept of permutationally invariant functions to represent non-smooth adiabatic potential energy surfaces as learnable smooth functions. Additionally, we leverage the foundational models developed in the Message Passing Atomic Cluster Expansion (MACE) \cite{batatia2022mace, batatia2023foundation, kovacs2023mace}, an architecture that has demonstrated exceptional performance in ground-state molecular systems by effectively accounting for higher-order atomic interactions and many-body effects, to transfer learn to excited states. It can be shown that, despite the distinct nature of different excited-state potentials, leveraging information gained from the ground state leads to a reduced amount of data needed to train excited-state potentials and improved generalization to unseen systems.

\section{Results}

%\cco{a lot of material in this section I would put under "methods" or maybe better "Theory". I would split off the numerical tests into a "Results" section. What is in the current methods section to my mind is a supplement or appendix.}

The proposed architecture, which we call X-MACE, extends the MACE machine learning interatomic potential model~\cite{batatia2022mace} to enable efficient parametrization of multiple excited states. Interpreting multiple energy levels as a “set,” we combine MACE with a DeepSet embedding into a latent space and an auto-encoder architecture to recover the non-smooth energy levels. The X-MACE architecture is illustrated in Figure~\ref{fig:method}a. Each component of the architecture is explained in detail in the remainder of this section.

% OLD TEXT
% An overview of The proposed X-MACE architecture is illustrated in Figure \ref{fig:method}a. As can be seen, the method introduces an autoencoder that is integrated into the framework to accurately model non-smooth excited-state PESs. Each part of the method will be explained in details in the remainder of this section. 

% OLD TEXT: 
% the numerical instability in computing their eigenvalues. This instability manifests as a non-linear relationship between the prediction error of the ESP coefficients and the resultant errors in the associated energy values. In practice, a prediction error of $\epsilon$ in the ESP coefficients can lead to significantly larger errors in the solutions. To address this issue, well-behaved polynomial families such as Chebyshev polynomials \cite{gutleb2024parameterizing} have been proposed to reduce such errors. However, the optimal polynomial set for a particular problem is likely to be dataset-specific and may not be universally applicable. Given these considerations, a promising approach is to leverage an autoencoder framework to learn a set of permutationally invariant functions. 

%\cco{a lot of material in this section I would put under "methods" or maybe better "Theory". I would split off the numerical tests into a "Results" section. What is in the current methods section to my mind is a supplement or appendix.}

\subsection{Learning non-smooth adiabatic excited states via sets of permutationally invariant functions}
\begin{figure*}[htbp]
    \centering
    \includegraphics[width=\textwidth]{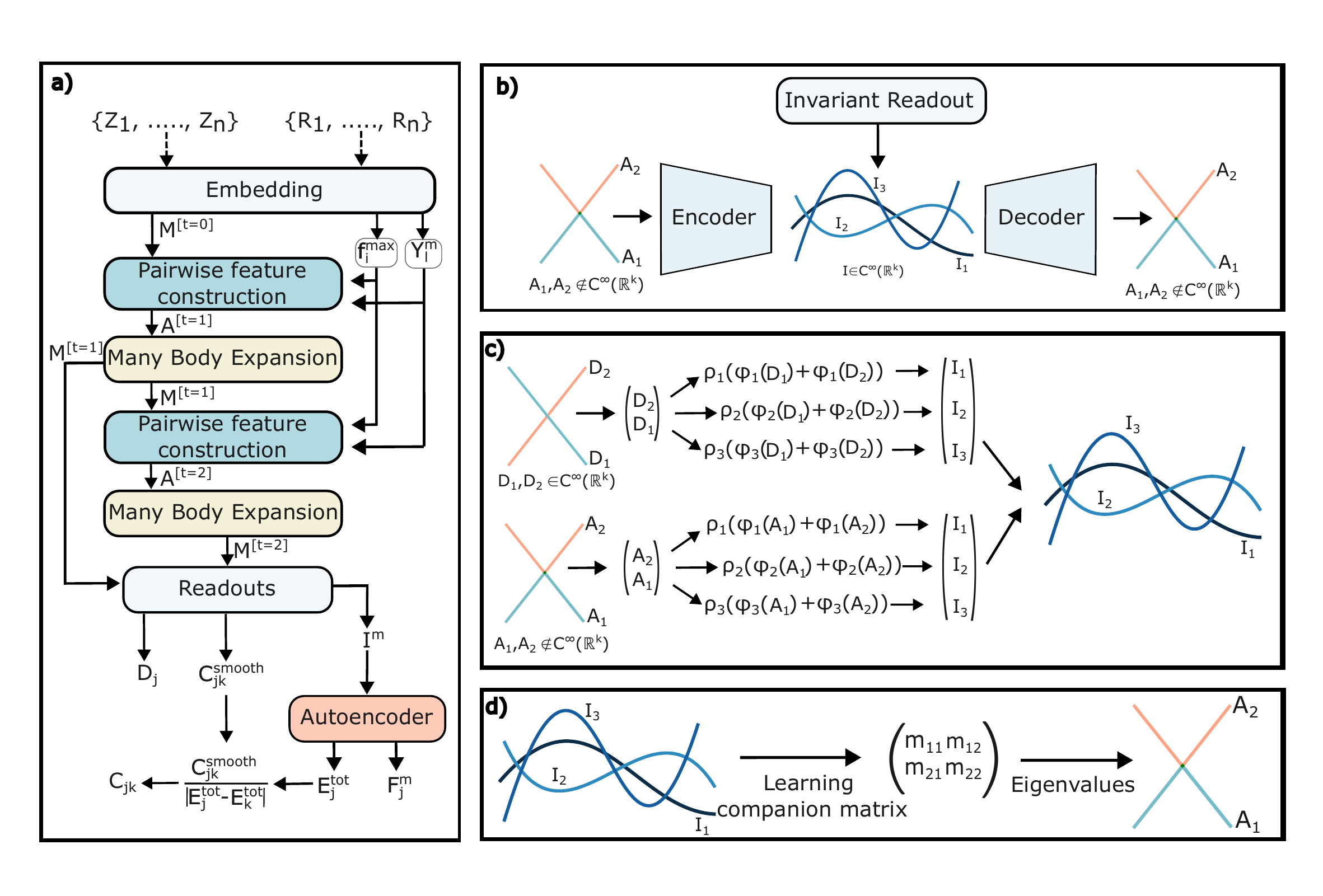}  % Replace with your image file
     \caption{\textbf{Architecture of the message-passing neural network (MPNN)-based Autoencoder Framework.} a) The model architecture consists of various iterations of MACE blocks consisting of single atom functions followed by many body expansions. The hidden states are passed to the readout functions; these are fitted to the learned permutationally invariant functions.  b) The output of the readout function in the MACE block is fitted to the latent space containing the permutationally invariant functions produced by the encoder. The corresponding permutationally invariant functions are mapped back to the energy surfaces using the decoder. Note that the $A_i$ correspond to the adiabatic or energetically sorted surfaces, $E_i$ in the paper.
     c) The encoder consists of a DeepSets architecture producing a series of permutationally invariant coefficients from the input energy levels. The coefficients are learned by the MPNN through the readout block. 
     d) The decoder reconstructs the original energy levels by learning a Hermitian companion matrix, ensuring real-valued eigenvalues.
     }
    \label{fig:method}
\end{figure*}
In ground-state dynamics, potential energy surfaces typically vary smoothly with the configuration space. By contrast, excited-state potential energy surfaces can exhibit cusps and non-smooth regions, so different representations are commonly used.\cite{Mai2015IJQC} The two most widely used representations are called the \emph{adiabatic} or \emph{diabatic} frameworks. The adiabatic PESs are constructed by evaluating the eigenvalues of the electronic Hamiltonian at each molecular conformation, providing an energetically sorted set of electronic states. While adiabatic surfaces are the direct output of quantum chemistry programs, they are most commonly used to propagate dynamics. However, when fitting these potentials with machine learning, they present significant challenges due to the presence of conical intersections. These regions lead to the formation of non-smooth hypersurfaces, which are problematic for smooth regressors, including neural networks. An example of adiabatic surfaces including such a conical intersection is shown in Figure~\ref{fig:method}b (left and right) with the two different adiabatic surfaces called $A_1$ and $A_2$.

Diabatic potential energy surfaces, on the other hand, are considered smoother functions of molecular geometry and are thus more suited to machine learning (Figure~\ref{fig:method}c, top-left). However, they are not straightforward to compute and not unique as they depend on a chosen reference point and are therefore often termed quasi-diabatic.%reordering states consistently across high-dimensional configuration spaces is difficult. 
Moreover, global unitary transformations can become extremely complex for large systems. Despite these computational challenges, (quasi-)diabatic surfaces are appealing for machine learning methods, such as sparse polynomials, tensor networks, or neural networks, because of their inherent smoothness.

Although adiabatic surfaces are not inherently smooth, they can be related to quasi-diabatic surfaces through sorting, and this relationship can be leveraged using permutationally invariant representations of the set of surfaces, e.g., elementary symmetric polynomials (ESPs) \cite{wang2023machine,gutleb2024parameterizing}, 
% . These polynomials, which are invariant under any permutation of their variables, are defined as follows:
%
\begin{equation}
\begin{aligned} 
e_1 = \sum_{i=1}^{n} E_i, \quad 
e_2 = \sum_{1 \leq i < j \leq n} E_i E_j, \\ e_k = \sum_{1 \leq i_1 < i_2 < \dots < i_k \leq n} E_{i_1} E_{i_2} \cdots E_{i_k}.
\end{aligned}
\label{eq:elementary_sym}
\end{equation}
%
%Due to the permutational invariance of these polynomials they
%\cco{I don't think this is the issue; the point is that they are smooth functions of the structure even in the presence of crossings or conical intersections.}
The ESPs are smooth functions of the adiabatic energies. Therefore, their values can be predicted efficiently using machine learning methods. The associated energy values can then be recovered numerically through a variety of methods, one notable example being the companion matrix, a matrix computed from the ESPs whose eigenvalues are the sought energy values.

A key challenge associated with using ESPs is that small errors in the fitted ESP values result in large errors in the reconstructed surfaces near crossings. 
For example, usage of the Frobenius companion matrix \cite{horn2012matrix, eastman2014companion} results in complex avoided crossings, while usage of the symmetric Schmeisser companion matrix \cite{schmeisser1993real} results in numerical instabilities. Another significant disadvantage is that despite ESPs being smooth, they are polynomials of the same degree as the number of surfaces which makes them challenging to approximate for increasing number of surfaces. 
Several variations of the idea are possible to ameliorate some of these challenges, \textit{e.g.}, employing the so-called Colleague matrix \cite{good1961colleague, trefethen2019approximation}, but an optimal approach is likely to be dataset-specific rather than universally applicable. 
Given these considerations, we propose a novel deep set autoencoder framework to learn a set of permutationally invariant functions.

\subsection{Encoding adiabatic energies: Leveraging DeepSets for Permutationally Invariant Functions}
%
%
%\cco{given the target audience for this paper I think it would be less confusing and more efficient to write this immediately in terms of energy levels rather than abstract ${\bf x}$. I would probably also call the latent space vector $\mathcal{I}$ or for "invariant", which also matches the figure. Also $F = f(E)$ would work.}
%
To learn non-smooth excited-state energy surfaces in a permutation-invariant manner, we employ an autoencoder architecture that operates directly on sets of energy levels. Let $\mathbf{E} = (E_1, E_2, \dots, E_n)$ be a vector of $n$ energy levels. The autoencoder consists of an encoder $f$ that maps $\mathbf{E}$ into a permutationally invariant vector $\mathcal{I}$, and a decoder $g$ that reconstructs the original energy levels:

\begin{equation}
    \tilde{\mathbf{E}} = g\bigl(\mathcal{I}\bigr) = g\bigl(f(\mathbf{E})\bigr),
\end{equation}

where $\tilde{\mathbf{E}} = (\tilde{E}_1, \tilde{E}_2, \dots, \tilde{E}_n)$ is the reconstructed output. The encoder and decoder are trained to minimize the difference between $\mathbf{E}$ and $\tilde{\mathbf{E}}$.
An illustration of the autoencoder can be seen in Figure~\ref{fig:method}b. 
Since the order of the $n$ energy levels is irrelevant, the encoder must be invariant to permutations. To maintain the permutational invariance, we employ a so-called deep set approach. A function $f(\mathbf{E})$ is permutationally invariant \cite{zaheer2017deep} if and only if it can be expressed as: 
\begin{align} f(\mathbf{E}) = \rho\left( \sum_{i=1}^n \phi(E_i) \right), 
\end{align}

where $\phi: \mathbb{R} \rightarrow \mathbb{R}^p$ maps individual elements to a higher-dimensional space, and $\rho: \mathbb{R}^p \rightarrow \mathbb{R}^k$ maps the aggregated result to the invariant vector $\mathcal{I}$. The sum operation ensures permutation invariance of the latent space representation $\mathcal{I} = f(\mathbf{E})$. The functions $\phi, \rho$ are chosen to be multi-layer perceptrons. This is illustrated in \ref{fig:method} c. 

While this approach ensures that the encoder is permutationally invariant, reconstructing the original energy levels from $\mathcal{I}$ poses a challenge. Since the functions $\phi$ and $\rho$ are modeled by large neural networks, solving them numerically to recover the individual energy levels is computationally intensive. To address this, we introduce a decoder architecture motivated by the companion matrix approach used to invert an ESP encoder. Our decoder first transforms the invariant $\mathcal{I}$ using a multi-layer perceptron (MLP) into a hermitian (self-adjoint) matrix $M$. The reconstructed energy levels $\tilde{\mathbf{E}}$ are then obtained by computing the eigenvalues of $M$. Constructing $M$ to be hermitian ensures that its eigenvalues (\textit{i.e.}, the reconstructed energy levels) are real. This is illustrated in \ref{fig:method} d. In general, the autoencoder framework can be expressed as:

\begin{align}
    M &= \phi\left(\rho\left( \sum_{i=1}^n \phi(E_i) \right)\right), \\
    \tilde{\mathbf{E}} &= \mathrm{Eig}(M).
\end{align}
where $\phi$ and $\rho$ are MLPs, and $M$ is the constructed Hermitian matrix. To map molecular structures to $\mathcal{I}$, we use the MACE architecture; however, it is worth noting that other architectures designed for molecular systems can also be employed here. A schematic overview of the MACE architecture and the process of fitting the invariants is shown in Figures~\ref{fig:method}a and~\ref{fig:method}b. Importantly, $\phi$ and $\rho$ can be combined into a single MLP for simplicity. The latent invariants $\mathcal{I}$, which are learned by MACE, can be extracted at any stage within the combined MLP, provided the summation operation in the DeepSet architecture has already been performed. Details on the loss function, training protocols, and computational settings can be found in the methods section. %and supplementary information. - you don't mention training details in the method section...

% OLD TEXT
% To avoid the clamping effect associated with complex eigenvalues seen when using other types of matrices. Hermitian matrices are described by the property:  \begin{equation} M = \bar{M}^\dagger, \end{equation} The reconstructed energy levels $\tilde{E}$ are obtained by computing the eigenvalues of the Hermitian matrix. 

%To map molecular structures to $\mathcal{I}$, we use the MACE architecture; it is worth noting that other architectures designed for molecules can be used here. A schematic overview of the MACE architecture and invariants fitting is shown in Figures~\ref{fig:method}a and~\ref{fig:method}b. Details on the loss function, training protocols, and computational settings can be found in the methods section and supplementary information.

\subsection{Modeling Conical Intersections}
%
%\cco{+++ CONTINUE HERE +++}
%\cco{[1] Seems to me this is the first actual results section; could this become the results section and merged with III. Transfer learning? [2] I would not call is ``modeling non-smooth functions''; the method is specifically about crossings / intersections. General non-smooth functions cannot be parameterized with this approach.}
%
 To demonstrate a model’s capability to accurately capture conical intersections, we first show slices through certain reaction coordinates thatrepresent a controlled scenario.%with limited complexity, where the models are directly trained on that specific slice. 
 Initially, we train our models on a wide range of conformations for the methylenimmonium cation, CH$_2$NH$_2^+$,  and SO$_2$ both systems that were chosen for their steep gradients around the conical intersections. %In case of the methylenimmonium cation, a non-smooth conical intersection. 
 Figure ~\ref{fig:energy_results} shows the resulting potential energy surfaces for the standard and autoencoder MACE architectures when compared to the reference values. In both cases, a reduction in error is seen around the conical intersections when using the autoencoder framework. For SO$_2$ the autoencoder replicates the reference curves, which intersect, which the standard MACE architecture fails to replicate. Additionally, an improvement is seen also in case of the methylenimmonium cation, where the standard MACE model obtains an error gap around 0.501eV whereas the autoencoder obtains an error of 0.275eV leading to a nearly 50\% smaller error for energy gaps. 

\begin{figure}[htbp] 
    \centering
    \includegraphics{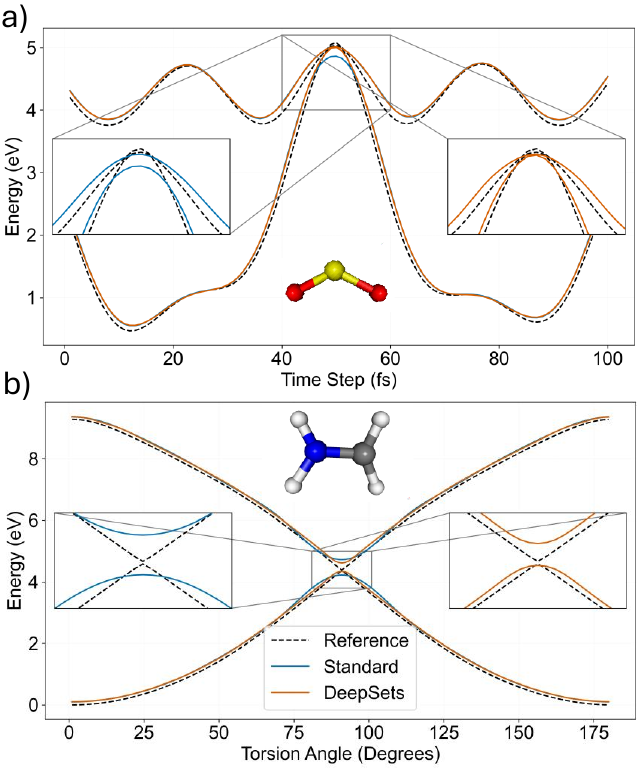}
    \caption{Figures showing predicted energy curves around conical intersections. a) First two potential energy surfaces of a trajectory of SO$_2$ through a conical intersection b) First two potential energy surfaces of a rotation around the torsion angle of the methylenimmonium cation}
    \label{fig:energy_results}
\end{figure}

To fully assess the robustness and generalization capability of our approach, we aim to extend the model to a specialized dataset enhanced with a number of conical intersections relevant for photochemical processes. We first start with investigation of the methylenimmonium cation and fulvene (C$_6$H$_6$). In case of the methylenimmonium cation, ultrafast transitions to the ground state take place upon light excitation to the second excited singlet state, S$_2$, due to isomerization, bipyramidalization, and rotation. Fulvene, in contrast, is an example of reflection, reflecting planar and twisted conical intersection geometries that lead to transitions between different potential energy surfaces. Thus, these tests involve training the model on a diverse set of molecular geometries and electronic states that encompass a wide range of molecular configurations over three and two electronic states, respectively. %As many geometries around conical intersections are included, these data sets represent good testbeds that are particularly difficult to model with conventional machine learning. 

\begin{figure*}[htbp] % Use figure* for a wide figure
    \centering
    \includegraphics[width=\textwidth]{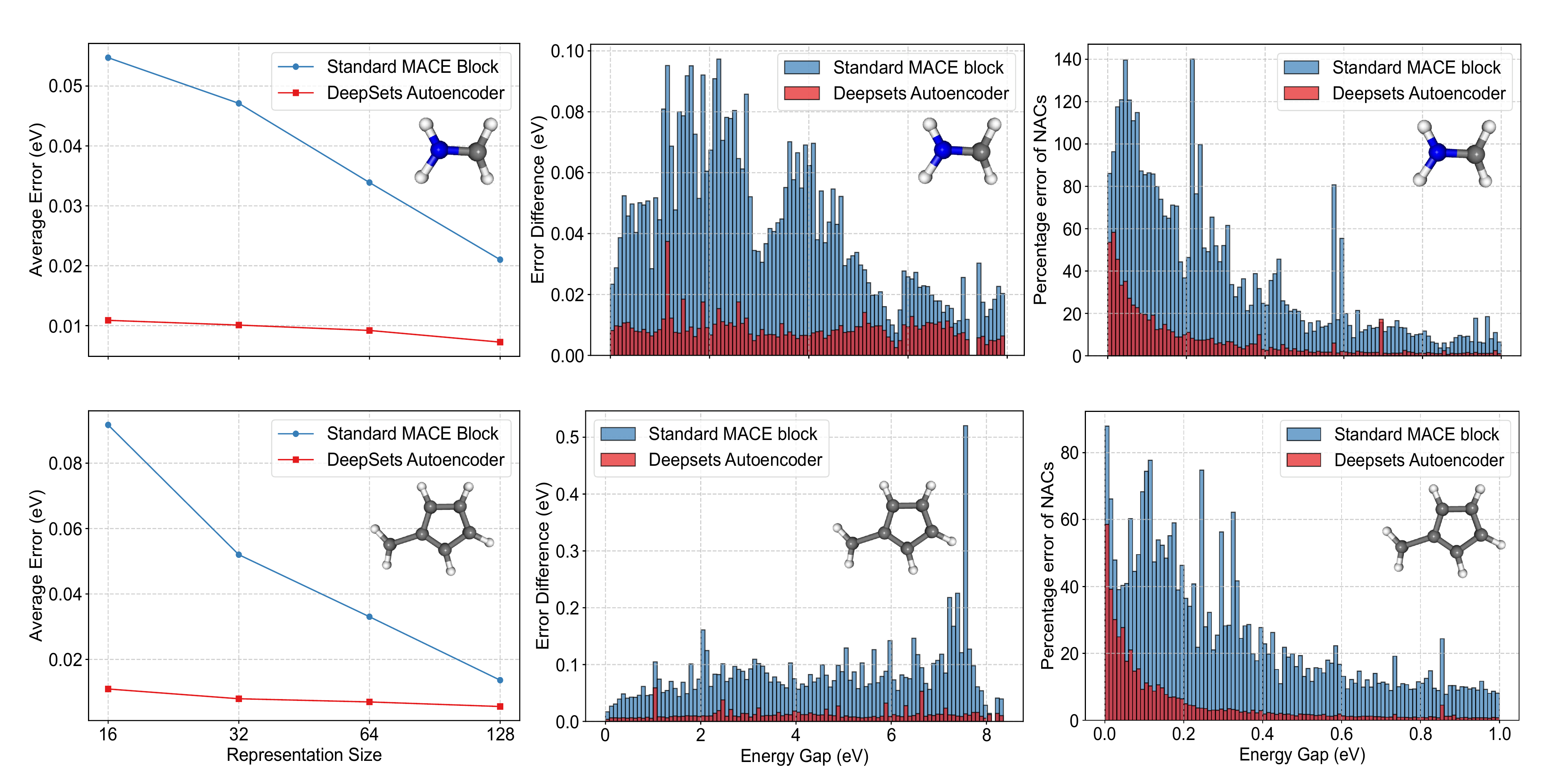}
    \caption{Energy-error dependence for the data set of a) the methylenimmonium cation and b) fulvene with varying representation sizes. A histogram plot of the energy-error values for c) the methylenimmonium cation and d) fulvene plotted against the energy gap between the potential energy surfaces. 
    Histogram plot showing the percentage error in the nonadiabatic coupling (NAC) values for e) the methylenimmonium cation and f) fulvene plotted against the energy gap.}
    \label{fig:energy_results}
\end{figure*}

In evaluating the model, we consider not only the energy distributions but also errors in nonadiabatic couplings that are defined as:

\begin{equation}
C_{mn} = \langle \psi_m | \nabla_\mathbf{R} \psi_n \rangle \approx
 \frac{\langle \psi_m | \nabla_\mathbf{R} H | \psi_n \rangle}{E_n - E_m},
 \label{eq:nonadiabatic}
\end{equation}

with $\psi_m$ being the wave function of state $m$. Using perturbation theory it can be shown that nonadiabatic couplings are approximately proportional to the inverse of the corresponding energy gap as shown in equation \ref{eq:nonadiabatic}. For a detailed derivation, see ref. \citenum{westermayr2020combining} for instance. As a consequence, singularities are present at conical intersections, as these mark regions in the energy landscape, where two potential energy surfaces become degenerate, and are thus especially difficult to model with neural networks. To avoid this issue, we multiply the nonadiabatic couplings with the energy gap to obtain smooth values (which we call "smooth NACs" in Table \ref{tab:Butene_Ethene_Propene_Methylammonium_metrics}), which can be modeled easier. However, the dependence on this inverse energy means that small errors in energies can lead to significant errors in nonadiabatic couplings around conical intersections. 

\begin{table*}[ht]
\centering
\begin{tabular}{llccc}
\toprule
\multirow{2}{*}{\textbf{Model}} & \multirow{2}{*}{\textbf{System}} & \multicolumn{3}{c}{\textbf{Mean Absolute Error (MAE)}} \\
\cmidrule(lr){3-5}
 &  & \textbf{Energy (eV)} & \textbf{Force (eV/\AA)} & \textbf{Smooth NACS (eV/\AA)} \\
\midrule
\multirow{4}{*}{SchNarc} 
& Chromophores      & 1.648 & 0.4422 & - \\
& Butene            & 0.07586 & 0.19102 & 0.14164 \\
& Ethene            & 0.05992 & 0.16944 & 0.19648 \\
& Methylenimmonium cation    & 0.1190 & 0.3800 & 0.4311 \\
\midrule
\multirow{4}{*}{SPaiNN} 
& Chromophores      & 0.5762 & 0.3852 & - \\
& Butene            & 0.0475 & 0.10285 & 0.135 \\
& Ethene            & 0.02793 & 0.08249 & 0.2108 \\
& Methylenimmonium cation    & 0.1018 & 0.3323 & 0.4231 \\
\midrule
\multirow{4}{*}{E-MACE} 
& Chromophores      & 0.13256 & 0.13790 & - \\
& Butene            & 0.02350 & 0.08859 & 0.10118 \\
& Ethene            & 0.0048 & 0.0359 & 0.1027 \\
&  Methylenimmonium cation     & 0.1295 & 0.2639 & 0.2540 \\
\midrule
\multirow{4}{*}{X-MACE} 
& Chromophores      & 0.10310 & 0.12602 & - \\
& Butene            & 0.01507 & 0.04169 & 0.2000 \\
& Ethene            & 0.0041 & 0.0311 & 0.10602 \\
& Methylenimmonium cation  & 0.0944 & 0.2309 & 0.2609 \\
\bottomrule
\end{tabular}
\caption{Mean Absolute Error (MAE) for energies, forces, and nonadiabatic couplings (NACS) averaged over all states of different models trained on excited states of butene, ethene, propene, and the methylenimmonium cation systems.}
\label{tab:Butene_Ethene_Propene_Methylammonium_metrics}
\end{table*}

%and how energy errors vary with representation size. 
Figures \ref{fig:energy_results} a and b show the decrease in energy-errors when adjusting the number of parameters or increasing the size of the representation in the graph neural networks. As can be seen, employing the autoencoder (red lines in the figure) significantly reduces errors in energies, even for models with smaller representation sizes. Specifically, the error for a representation size of 128 in the excited-state adapted standard MACE block (X-MACE) remains higher than that of the autoencoder model with a size of just 16. This improvement is likely due to the autoencoders ability to construct smooth functions from the potential energy surfaces. The constructed smooth surfaces have a lower complexity. Consequently,  substantially fewer parameters are required in the X-MACE block.

Moreover, the autoencoder markedly narrows the distribution of errors in energy predictions across the whole dataset compared to standard X-MACE (\textit{i.e.}, without the autoencoder), as shown in \ref{fig:energy_results} c and d. The errors were generated by averaging the prediction values across four models with varying representation sizes. It can be seen that the errors are reduced for energies of the methylenimmonium cation as well as of fulvene across the whole energy distribution. This is again likely due to the lower complexity functions needed to model the smooth invariant surfaces constructed using the autoencoder (red blocks). %In addition, the reduced pollution caused by the conical intersections in other regions of the fitted potential energy surfaces can lead to this result.  

While the reduced error in the energies is advantageous, its primary significance lies in the corresponding decrease in the error of the nonadiabatic couplings. As already briefly mentioned above, nonadiabatic couplings are important in photodynamics simulations using, \textit{e.g.}, the surface hopping method, to determine the probability of transitions between different potential energy surfaces. Due to the inverse proportionality of the nonadiabatic couplings to the energy gap \ref{eq:nonadiabatic}, the transition probability is often large close to conical intersections, while small elsewhere. Thus, it can make a significant difference to the simulation if the nonadiabatic couplings are modeled inaccurately in these regions. For both systems, the methylenimmonium cation and fulvene, it can be seen that the autoencoder drastically outperforms the standard MACE block \ref{fig:energy_results}. However, both models exhibit larger errors in the proximity of conical intersections. This occurs because near-degenerate eigenvalues remain inherently unstable, resulting in larger errors.

To demonstrate the performance of X-MACE both with and without the autoencoder framework against existing models we compared it against two established excited-state models: SchNarc\cite{westermayr2020combining} and SPaiNN,\cite{mausenberger2024spainn} which are based on the SchNet\cite{schutt2018schnet} and PaiNN\cite{schutt2021equivariant} representations, respectively. Results are shown in table \ref{tab:Butene_Ethene_Propene_Methylammonium_metrics}. The evaluation was conducted using four datasets: three represent alkenes (ethene, propene, and butene), taken from ref. \citenum{mausenberger2024spainn} (for details see the Methods section), and the fourth encompasses a diverse set of chromophore systems, which were generated in this work (see Methods section for details).
For all architectures, we trained models across a range of representation sizes and averaged the results. Additional details how the errors vary with varying representation sizes  are provided in the supplementary material in section S3 in Tables S1-S4.
As can be seen, X-MACE outperforms SchNarc and SPaiNN in predicting energies, forces, and outperforms or performs equally in cases of the smooth nonadiabatic couplings. When comparing X-MACE with and without the autoencoder framework, the errors in smooth nonadiabatic couplings remain somewhat similar, as expected, since the readout block is not modified. However, notable reductions in errors are observed for energies and forces, particularly for energies albeit to a lesser extent than those seen in results represented in Figure \ref{fig:energy_results}, as the respective datasets were supplemented with a large number of geometries representing conical intersections. The data for butene, ethene, and the methylenimmonium cation represent different conformations of the same molecule. The chromophores data set created in this work contains different conformers of different molecules. As is visible from the table, the improvement between the X-MACE variants compared to SchNarc and SPaiNN are substantial.For energy values, the X-MACE variants produced errors more than an order of magnitude smaller than SchNarc and approximately one-third of the errors observed with SPaiNN. A similar observation can be seen for the forces with substantial improvements when using X-MACE.%Additionally, we constructed a dataset consisting of a wide range of chromophore systems computed at an ADC(2) level, in this case the improvement between the X-MACE variants compared to SchNarc and SPaiNN was substantial. For energy values the X-MACE variants obtained errors of over an order of magnitude lower than SchNarc and a third of the errors obtained by SPaiNN. A similar observation can be seen for the forces where substantial improvements were seen.

\subsection{Transfer Learning from Ground State to Excited State}
%
%\cco{to my mind this can just become two additional results sub-sections. Maybe it can be compressed a little bit but overall seems fine as is. Overall this seems fine, but I plan to re-read and potentially edit "Generalizations between systems" one more time. This is still a puzzle for me. }
%

\begin{figure*}[htbp] % Use figure* for a wide figure
    \centering
    \includegraphics[width=\textwidth]{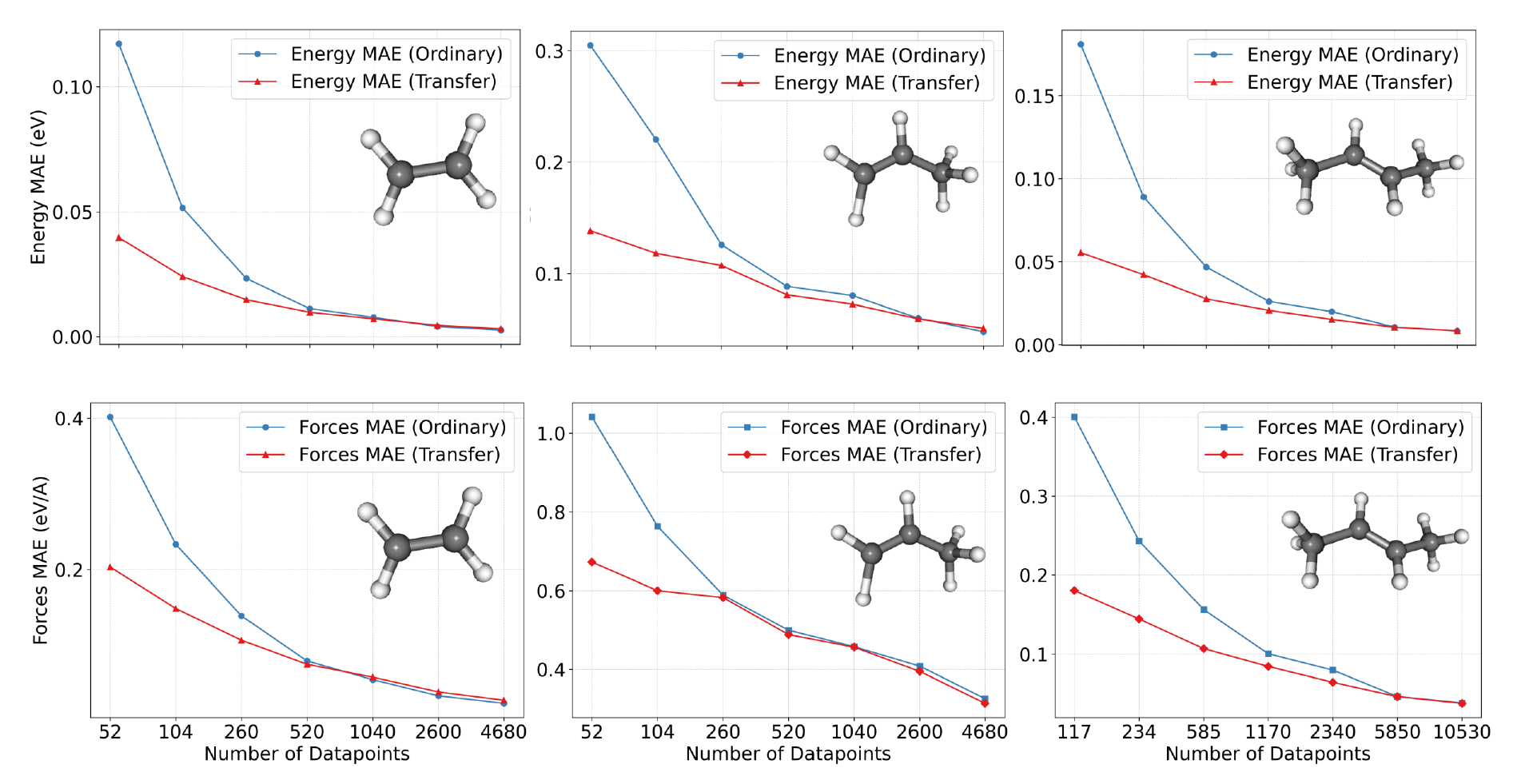}
     \caption{Learning curves showing the variation of the error in the energies (first line) and forces (second line) with increasing amount of data for ethene (a,d), propene (b,e), and butene (c,f).}
    \label{fig:transfer_results}
\end{figure*}

Foundational machine learning models have transformed numerous scientific domains by offering robust, versatile frameworks capable of capturing intricate patterns and generalizing across diverse datasets.\cite{reviewgeneral} Although these models have already accelerated progress in ground-state simulations, comprehensive counterparts for excited-state systems remain elusive. Adapting machine learning force fields from ground to excited states is particularly challenging due to the complexity of excited-state phenomena and the different nature of the different electronic states. Factors such as nonadiabatic transitions, sharp discontinuities, and non-smooth regions in potential energy surfaces make excited states inherently more difficult to model than their ground-state counterparts. Moreover, the limited availability of excited-state data stemming from the demanding quantum chemical calculations required further complicates the training of robust, generalizable models. Developing methods capable of transferring knowledge within excited states would be a significant breakthrough, as it is commonly believed to be infeasible due to the inherent differences between electronic states, with many considering the creation of transferable excited-state potentials impossible.\cite{Westermayr2021CR,reviewgeneral} 

Here, we present initial efforts to leverage foundational machine learning models as a baseline for transferring to excited states by fine-tuning their internal representations using only a small amount of excited-state data. While accurately capturing the non-smooth features of excited-state potential energy surfaces without explicit localization within conformational space is difficult,  our approach suggests that it may be possible to extrapolate useful information for excited states by using the autoencoder and refining representations learned from ground-state scenarios.
%Transferring machine learning force fields from ground-state to excited-state presents significant challenges due to the intricate nature of excited-state phenomena. The presence of nonadiabatic transitions, along with sharp discontinuities and non-smooth regions in the potential energy surfaces (PES), makes excited states particularly difficult to model using ground-state data. Moreover, the scarcity of high-quality excited-state data—stemming from the computational intensity of the required quantum chemical calculations limits the ability to train robust and generalizable models. 
%However, while modeling the non-smooth components of the PES without explicit knowledge of their localization in conformational space is challenging, it may be possible to extrapolate knowledge of the excited-state PES through fine-tuning the internal representations of existing ground-state models using far less data.

%Many studies have demonstrated the effectiveness of transfer learning, and more recently, it has proven effective in capturing differences across various levels of quantum theory through delta learning or by reusing the internal representations of existing machine learning models to reduce the amount of data needed. 
One particular advancement in recent years are the MACE foundational models;  \cite{batatia2022mace, batatia2023foundation, kovacs2023mace, smith2017ani, deng2023chgnet, gao2020torchani} these models have been trained on large datasets to predict the properties of a vast number of systems covering most of the periodic table at near density functional theory accuracy. Fine-tuning these models involves re-optimizing their parameters using a new, smaller dataset. In the case of excited states, we initially retain the internal representations of the MACE foundational models and randomly initialize the readout functions. The parameters obtained are used as an initial starting point for the training of the energies or permutationally invariant function values of the smaller dataset. In our case, we refine the representations for X-MACE starting from the foundational MACE-OFF representation. %However only fine tuning the readouts and holding the representation fixed lead to a larger error so will be left to the supplementary material.

In figure \ref{fig:transfer_results} the error curves for energies and forces with different percentages of data are compared for three systems: ethene, propene and butene.% with data sets comprised from photodynamics of ref. \cite{mausenberger2024spainn}. 
All three systems were trained using the MACE architecture fitted with the autoencoder framework. Consistent for all cases can be seen that with a lower amount of data, between 1\% to 10\% of the full data set, which are smaller than 100 data points in total, the transfer learning outperforms the standard learning process and particularly excels when the amount of data is very small. However, when the amount of data increases, the difference between the transfer and ordinary learning becomes negligible. Thus for the cases where data is highly expensive or scarce, which is the case for most excited-state systems, transfer learning starting from the ground state will lead to a substantial error decrease and thus expected improved stability of dynamics simulations.

\subsection{Generalization Outside of Dataset}
In addition to the transferability from ground state information to excited states, the transferability of excited-state properties in chemical space has shown to be an open challenge.\cite{Axelrod2022NC,Westermayr2020JCP} Especially modeling larger molecular systems presents a significant challenge due to the exponential increase in configurational space. As molecular size increases, so does the number of possible conformations and accessible electronic states within a certain energy window, resulting in comparably more critical regions on potential energy surfaces. In addition, the larger the flexibility of a molecular system, the larger the conformational space that needs to be sampled to cover all possible photo-deactivation processes.

\begin{figure*}[htbp] % Use figure* for a wide figure
    \centering
    \includegraphics[width=\textwidth]{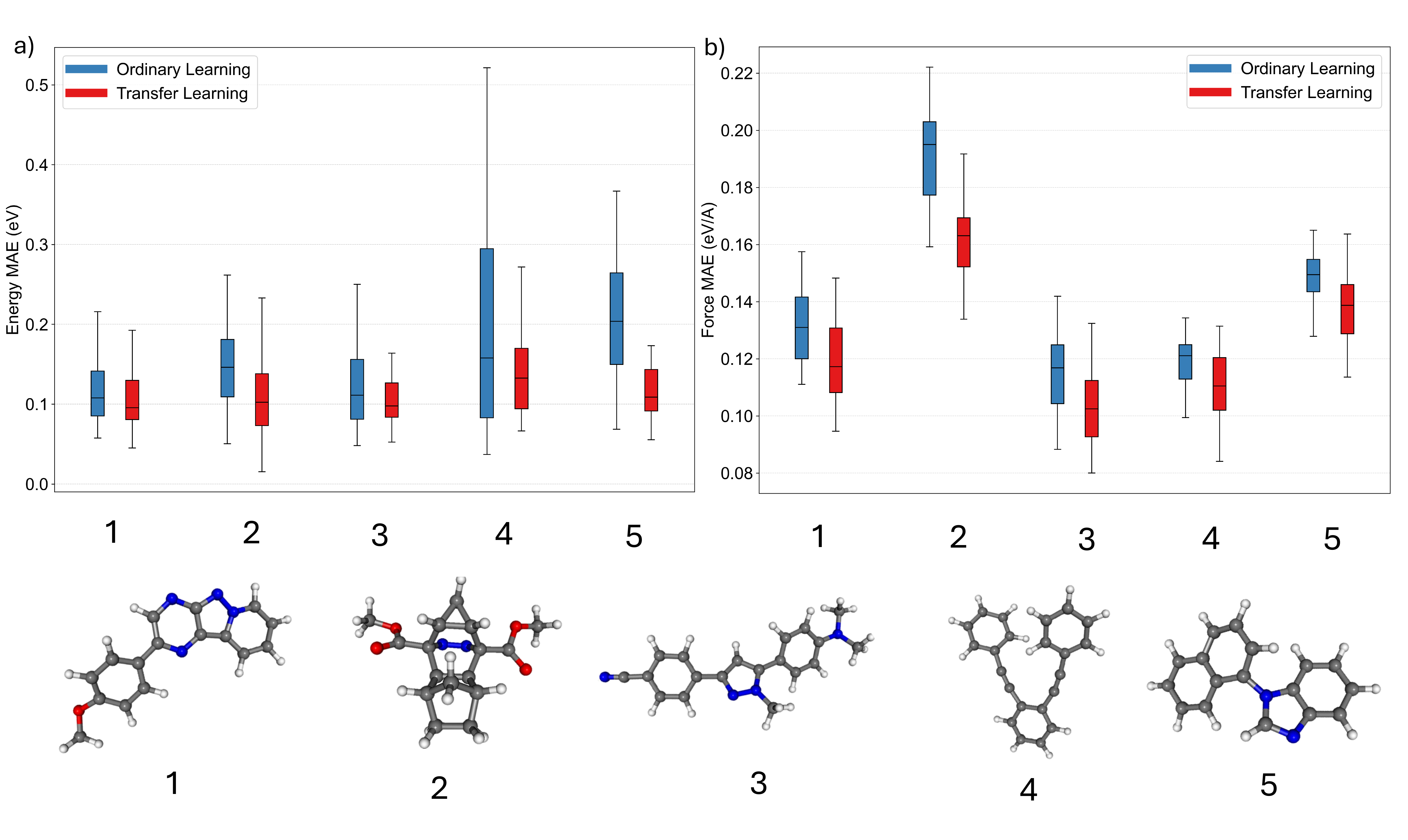}
    \caption{Boxplots representing the distribution of a) energy and b) force errors for five systems outside the training set that can be seen below the plots. Two models are compared: transfer learning (red) and training from scratch (blue).}
    \label{fig:general}
\end{figure*}
Finally, to facilitate the transferable prediction of excited states over a wide range of chemical systems, we have constructed a dataset consisting of several chromophore systems, details can be found in the methods section. The construction of the data set is specified in the method section. A total amount of 12183 data points was computed using %As a reference method, we have chosen the Algebraic Diagrammatic Construction method, ADC(2), level for the first five singlet states of each system. 
chromophores that are of particular interest to photochemical processes due to their presence in biological systems and their ability to absorb light at energy windows within the UV/Visible range. To investigate the prediction accuracy of models trained on this data set, we investigate the range of errors for a series of unseen chromophores and compare the errors between learning from randomly initialized parameters (ordinary learning) and transfer learning models for energies and forces. As before, transfer learning models are based on the foundational MACE-OFF model. The results can be seen in Figure \ref{fig:general} and are illustrated using boxplots. In both cases, it can be seen that the models generalize well since training errors are within a range of around 0.1eV for energies and 0.1eV/{\AA} for forces. While these errors might seem large compared to ground-state systems, these energy errors are comparable or only slightly larger than typical excited-state ML models.\cite{westermayr2021physically, ghosh2019deep, stuke2019chemical,westermayr2019machine,westermayr2020combining,mausenberger2024spainn,westermayr_2022_deep} However, for the five unseen systems the average error is smaller, when knowledge is transferred from the ground-state. Surprisingly, the error is distributed almost equally between the different states, even though the transfer learning model stems from ground-state data only. The errors for each state separately are plotted in the supporting information in Figures S1-S6. This reduction in error is likely due to internal geometrical objects inside the representation giving a unique description for each molecular system which is then extrapolated to predict excited-state properties. %However, keeping the representations fixed while only optimizing the readout functions results in higher errors compared to both full transfer learning and training the model from scratch. This increased error is likely because the internal representations require fine-tuning to accurately capture the essential geometric features of the excited-state potential energy surfaces.

\subsection{Summary}
In this work, we address the accurate modeling of non-smooth potential energy surfaces that are particularly relevant for photochemical processes and appear near conical intersections -- regions on potential energy landscapes that have a large influence of a reaction outcome. Conventional neural network approaches encounter difficulties in these regions due to their intrinsic non-smoothness and complexity. In this work, we develop an excited-state version of one of the most popular ground-state message-passing neural network models, \textit{i.e.}, MACE (message passing atomic cluster expansion), which we term  X-MACE.%While previous strategies have employed specific families of permutationally invariant polynomials, the optimal choice of polynomial basis generally depends on the system under investigation.
Our methodology improves upon previous excited-state models by integrating learnable permutationally invariant functions through an autoencoder architecture. Using this framework, we demonstrate that our approach not only substantially outperforms direct energy prediction methods, but also enables transferable predictions across different molecules in their excited states. Especially in the vicinity of conical intersections, large improvements can be achieved in for energy predictions and other excited-state properties like couplings using the X-MACE framework. The inclusion of the autoencoder enhances accuracy across the conformational space while simultaneously reducing the number of parameters needed to fit excited-state potential energy surfaces. This efficiency likely arises from the autoencoder’s ability to construct a smoother, lower-complexity representation thereof.

Another notable outcome of this work is the demonstration of robust transferability from ground-state machine learning models to excited states. By fine-tuning pre-trained foundational MACE models, we significantly diminish the amount of excited-state data required for training. Starting from a pre-trained model, as little as 1\% of the data typically required for accurate excited-state machine learning potentials can be used, resulting in performance that exceeds models trained from scratch with randomly initialized parameters. This is particularly beneficial given the computational expenses associated with generating excited-state data. Moreover, this transferability extends beyond the initial training set, reducing errors by approximately 30\% for energies and 15\% for forces when applied to chromophores outside the original training domain. Notably, the maximum error is also reduced across most states, narrowing the energy error window and likely improving the stability of photodynamics simulations.

%Future work will look at incorporating X-MACE and the  autoencoder framework in surface hopping simulations to evaluate its performace in dynamics simulations. 
Future work will further explore the propensity of foundational excited-state models, paving the way towards photochemical studies that are currently not feasible due to the high computational efforts required to compute accurate excited-state data.

\section{Methods}
\subsection{Quantum Chemical Reference Calculations and Dataset Generation}
The SO$_2$ dataset was taken from \cite{westermayr2020combining} and generated using the linear vibronic coupling model\cite{plasser2019lvc} and where more information about the dataset generation can be found. The methylenimmonium cation dataset, without added conical intersections, was computed for three states using the multi-reference configuration interaction single double (MR-CISD) method using Columbus \cite{Lischka2011columbus} based on state-averaged complete active space self-consistent field (SA-CASSCF) using 6 active electrons in 4 active orbitals averaged over 3 states (SA(3)-CASSCF(6,4)) with the basis set aug-cc-pVDZ. The dataset was taken from \cite{westermayr2019machine}. 
The methylenimmonium cation dataset, with added conical intersections was computed using the same method, \textit{i.e.}, MR-CISD(6,4) with the basis set aug-cc-pVDZ. The structures in the dataset consist of the previous dataset combined with conical intersections optimized using the SHARC interface.
The fulvene dataset was computed using SA(2)-CASSCF(6,6) method averaged over two states using 6 active electrons in 6 active orbitals. This dataset will be published in an additional publication at a later date and is available upon request.

The ethene, propene and butene datasets were taken from \cite{mausenberger2024spainn} and data is based on the CASSCF method using the double-$\zeta$ cc-pVDZ basis set. The data set contains three singlet states and full details can be found in the respective publication.

The chromophores dataset was generated with this study. Therefore,  a total of 368 SMILES strings were taken from \cite{joung2020experimental}. Initial geometries were constructed using RDKit \cite{landrum2013rdkit} and then expanded as follows: Metadynamics simulations were performed using the xtb program \cite{bannwarth2019gfn2},%cite xtb and program - orca (Nope ase)?
with a bias potential strength of 0.1405 Hartree (kpush) and a Gaussian width of 0.01125 (alp). The underlying molecular dynamics simulations were conducted at 300 K with a timestep of 1.0 fs, a total simulation time of 10 ps, and trajectory data recorded every 200 fs. The recorded data points were recomputed with the Algebraic Diagrammatic Construction Scheme to second order perturbation theory (ADC(2)) \cite{Dreuw2015ADC2} for the first 5 singlet states. %Calculations which failed were discarded. 
A total of 12183 calculations converged, resulting in our final data set.

\subsection{MACE}
Message Passing Neural Networks (MPNNs) are a class of neural networks tailored for graph-structured data. They have gained prominence in molecular modeling due to their ability to capture both local and global structural information through iterative updates of node features. In molecular systems, nodes typically represent atoms, and edges represent chemical bonds or interactions between them.

Formally, a molecular system can be represented as a graph \( G = (V, E) \), where \( V \) is the set of nodes and \( E \) is the set of edges. Each node \( i \in V \) is associated with a feature tuple at layer \( t \):

\begin{align}
    \sigma_i^{(t)} = \left( \mathbf{h}_i^{(t)}, Z_i, \mathbf{R}_i \right),
\end{align}

where \( \mathbf{h}_i^{(t)} \) is the hidden feature vector of node \( i \) at iteration \( t \), \( Z_i \) is the atomic number of atom \( i \), and \( \mathbf{R}_i \) is the position vector of atom \( i \) in three-dimensional space.

The MPNN operates in two key stages: the message passing phase and the readout phase. During the message passing phase, node features are updated over \( n \) iterations to capture the structural information of the graph.

At each iteration \( t \), a message \( \mathbf{m}_i^{(t)} \) is constructed for each node \( i \) by aggregating information from its neighboring nodes \( \mathcal{N}(i) \):

\begin{align}
    \mathbf{m}_i^{(t)} = \sum_{j \in \mathcal{N}(i)} M\left( \mathbf{h}_i^{(t)}, \mathbf{h}_j^{(t)}, \mathbf{r}_{ij} \right),
\end{align}

where \( M \) is the message function that combines the features of node \( i \), its neighbor \( j \), and their relative position \( \mathbf{r}_{ij} = \mathbf{R}_j - \mathbf{R}_i \).

The hidden state of node \( i \) is then updated using an update function \( U_t \):

\begin{align}
    \mathbf{h}_i^{(t+1)} = U_t\left( \mathbf{h}_i^{(t)}, \mathbf{m}_i^{(t)} \right).
\end{align}

This iterative process allows the network to propagate information through the graph, enabling each node to accumulate knowledge about its local neighborhood.

To capture higher-order interactions beyond simple pairwise relationships, an atomic cluster expansion (ACE) \cite{dusson2022atomic, ho2024atomic} can be integrated into the MPNN framework. This expansion allows the model to consider many-body terms, which are crucial for accurately modeling complex molecular interactions. The integration of this with the message passing framework leads to the MACE model\cite{batatia2022mace}.

The message \( \mathbf{m}_i^{(t)} \) at iteration \( t \) is then expressed as a sum over many-body contributions:

\begin{align*}
    \mathbf{m}_i^{(t)} &= \sum_{j \in \mathcal{N}_1(i)} u_1 \left( \sigma_i^{(t)}, \sigma_j^{(t)} \right) \\
    &+ \sum_{j_1 \in \mathcal{N}_2(i)} \sum_{j_2 \in \mathcal{N}_2(j_1)} u_2 \left( \sigma_i^{(t)}, \sigma_{j_1}^{(t)}, \sigma_{j_2}^{(t)} \right) \\
    &\quad \vdots \\
    &+ \sum_{j_1, \dots, j_\nu \in \mathcal{N}_\nu(i)} u_\nu \left( \sigma_i^{(t)}, \sigma_{j_1}^{(t)}, \dots, \sigma_{j_\nu}^{(t)} \right),
\end{align*}

where \( u_\nu \) is the interaction function for the \( \nu \)-body term, and \( \mathcal{N}_\nu(i) \) denotes the set of \( \nu \)-th collection of neighbors of node \( i \).

An essential aspect of modeling physical systems is ensuring that the model's outputs transform correctly under rotations and translations. In this context, messages are constructed to be equivariant with respect to the rotation group \( SO(3) \). This means that if the atomic positions are rotated, the predicted properties, such as forces or energy, transform appropriately. Ensuring equivariance is critical for the physical accuracy of the model.\cite{white2021deep}

In the readout phase, the model aggregates information from the updated hidden states to compute global properties of the molecular system. Each node contributes to the final prediction, and their contributions are summed as follows:

\begin{align}
    \rho = \sum_{i \in V} R\left( \mathbf{h}_i^{(t)} \right),
\end{align}

where \( R \) is the readout function that maps the hidden state of node \( i \) to a scalar contribution, and \( \rho \) represents a property of the molecular system, such as total energy, dipole moment, or various coupling constants.

\subsection{Loss Function}

The training process optimizes a total loss function that combines two components: the reconstruction loss and the invariant vector matching loss. 

The reconstruction loss is based on the Mean Squared Error (MSE) between the true energy levels \( \mathbf{E} = (E_1, E_2, \dots, E_n) \) and their reconstructed counterparts \( \tilde{\mathbf{E}} = (\tilde{E}_1, \tilde{E}_2, \dots, \tilde{E}_n) \), given by:
\begin{equation}
    \mathcal{L}_{\text{reconstruction}} = \frac{1}{n} \sum_{i=1}^n \bigl(E_i - \tilde{E}_i\bigr)^2,
\end{equation}
where \( n \) is the number of energy levels in the dataset. This loss ensures that the autoencoder accurately reconstructs the input energy levels from the learned permutationally invariant representation.

To align the output of the MACE architecture with the permutationally invariant representation \( \mathcal{I} \) learned by the autoencoder, we define an additional matching loss:
\begin{equation}
    \mathcal{L}_{\text{matching}} = \frac{1}{N} \sum_{j=1}^N \bigl(\mathcal{I}_j^{(\text{MACE})} - \mathcal{I}_j^{(\text{AE})}\bigr)^2,
\end{equation}
where \( \mathcal{I}_j^{(\text{MACE})} \) is the \( j \)-th permutationally invariant vector predicted by MACE  and \( \mathcal{I}_j^{(\text{AE})} \) is the corresponding permutationally invariant vector obtained from the autoencoder. Here, \( N \) represents the total number of permutationally invariant vectors in the constructed.

However, the permutationally invariant space of the autoencoder can be highly non-linear. A fixed error in this representation does not necessarily translate into a consistent energy error, as the error can vary depending on the direction within the permutationally invariant space. To address this, we propagate the derivatives back through the autoencoder and MACE during training. This approach allows the model to adjust its parameters in a way that directly reduces reconstruction errors while accounting for the non-linear errors in the energy that are obtained due to the complexity of the invariant space.

The total loss function governing the model training is the sum of the reconstruction loss and the matching loss:
\begin{equation}
    \mathcal{L}_{\text{total}} = \mathcal{L}_{\text{reconstruction}} + \mathcal{L}_{\text{matching}}.
\end{equation}

During inference, only the decoder of the autoencoder is utilized. Given a new molecular geometry \( \mathcal{G} \), MACE predicts the permutationally invariant vector \( \mathcal{I} \), which is then mapped by the decoder \( g_\phi \) to the final reconstructed energy levels \( \tilde{\mathbf{E}} \).

\section{Training Details}
During all experiments, the model hyperparameters were kept constant for the MACE models except for the representation sizes. In these cases the number of channels used was 16, 32, 64, 128. A 5~{\AA} cutoff radius was applied to all models and 2 interaction layers were used in all cases. Spherical harmonics up to degree 3 were used and in all cases where the deep sets auto-encoder was used 16 learnable invariants were used. An additional loss function is added for the remaining properties. This loss is comprised of the MSE over all remaining properties, in this case forces and non-adiabatic couplings;

\begin{align}
L_{other} = 
&\; a_F \sum_{i=1}^n \bigl(F_{REF}^{i} - \hat{F}^{i}\bigr)^2 \nonumber \\
&+ a_C \sum_{k<i}^{n} \sum_{i=1}^n 
\begin{split}
\min \Bigl( &\bigl\| C_{REF}^{i,k} - \hat{C}^{i,k} \bigr\|^2, \bigl\| C_{REF}^{i,k} + \hat{C}^{i,k} \bigr\|^2 \Bigr)
\end{split}
\end{align}
where $a_F$ and $a_C$ were all set to 100 if $a_C$ was included during the training. Here $C^{(i,k)}$ corresponds to the nonadiabatic coupling between the states $i$ and $k$. In all cases, the default MACE optimizer was used with a learning rate of 0.001. As can be seen, we use a minimum function for fitting nonadiabatic couplings, which stems from the fact that nonadiabatic couplings are only defined up to an arbitrary sign due to the arbitrary phase of the wave function.\cite{westermayr2019machine,akimov2018JPCL} Thus, properties need to be trained phase-free. Details about the phase free training in MACE can be found in the supporting information in section S2 and additionally in ref. \citenum{westermayr2020combining}.  

The SPaiNN models using the SchNet and PaiNN representations were performed using a similar loss function. Here the loss used was a linear combination between $L_{other}$ and $L_{reconstruction}$. The associated weighting to each of the properties was equal during the training of these model. For all models two interaction layers were used for fair comparison with the proposed architectures and the number of channels in the representation was varied according to the experiment conducted. The learning rate used was $10^{-4}$ (default) and all other hyperparameters were the same as the defaults for these models.

 \section*{Acknowledgements}
This work is funded in parts by the Deutsche Forschungsgemeinschaft (DFG) -- Project-ID 443871192 - GRK 2721: "Hydrogen Isotopes $^{1,2,3}$H". The authors acknowledge the ZIH TU Dresden, the URZ Leipzig University and Paderborn Center for Parallel Computing (PC2) for providing the computational resources. 

 \section*{Author Contributions}
RB (Conceptualization, Data curation, Formal analysis, Investigation, Methodology, Software, Visualization, Validation, Writing - original draft). 
CO (Methodology, Writing - review).
JW (Methodology, Conceptualization, Supervision and Discussion, Writing - review).

\section*{Materials and Correspondence}
Correspondence to Julia Westermayr. 
\section*{Competing Interests}
The authors declare no competing interests.

\section*{Code Availability}
The code is publicly available at: \url{https://github.com/rhyan10/X-MACE}

\section*{Data Availability}
The data sets are available at the respective references mentioned in the text. The data generated for this work is available on figshare under: \url{https://figshare.com/articles/dataset/Datasets_for_X-MACE_Publication/28425173}

%%%END OF MAIN TEXT%%%



%%%REFERENCES%%%
\providecommand*{\mcitethebibliography}{\thebibliography}
\csname @ifundefined\endcsname{endmcitethebibliography}
{\let\endmcitethebibliography\endthebibliography}{}

\end{document}

\maketitle
\tableofcontents
\clearpage

\section{State Error Breakdown for Chromophores}
The following figures show the breakdown of mean absolute errors (MAEs) for energies and forces according to Figure 5 in the main paper. 

\begin{figure*}[htbp] % Use figure* for a wide figure
    \centering
    \includegraphics[width=\textwidth]{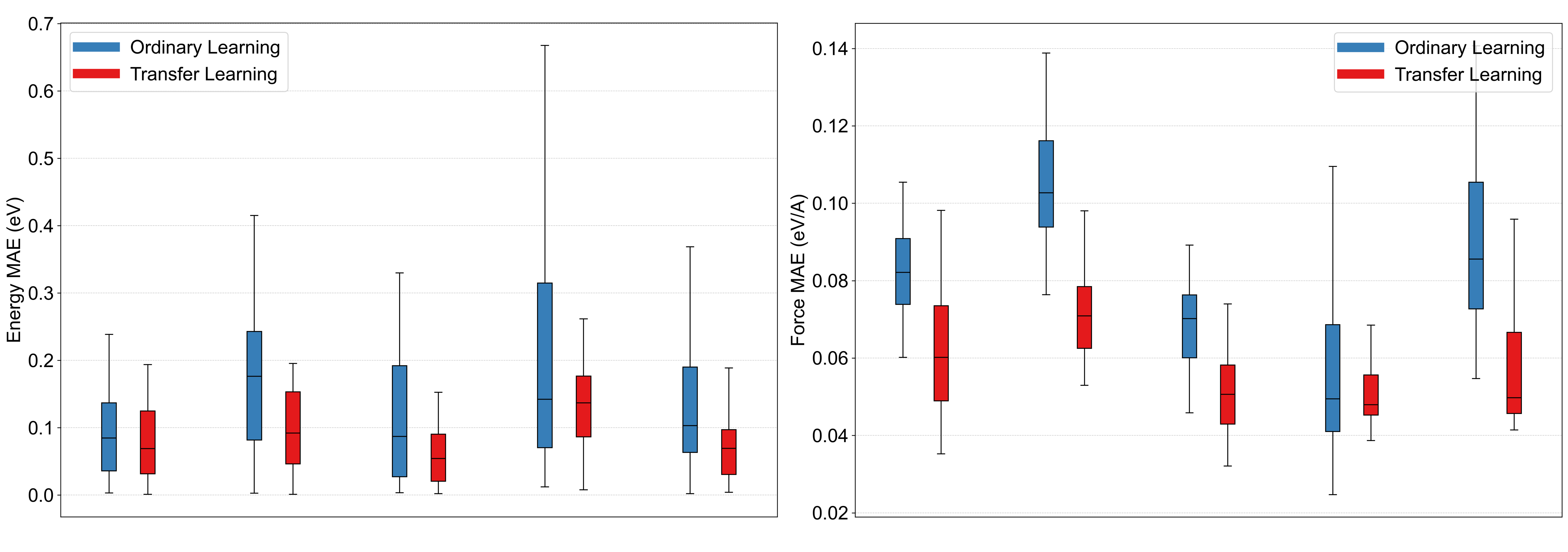}
    \caption{Boxplots representing the distribution of a) energy and b) force errors for five systems outside the training for the ground state. Two models are compared: transfer learning (red) and training from scratch (blue).}
    \label{fig:state1}
\end{figure*}

\begin{figure*}[htbp] % Use figure* for a wide figure
    \centering
    \includegraphics[width=\textwidth]{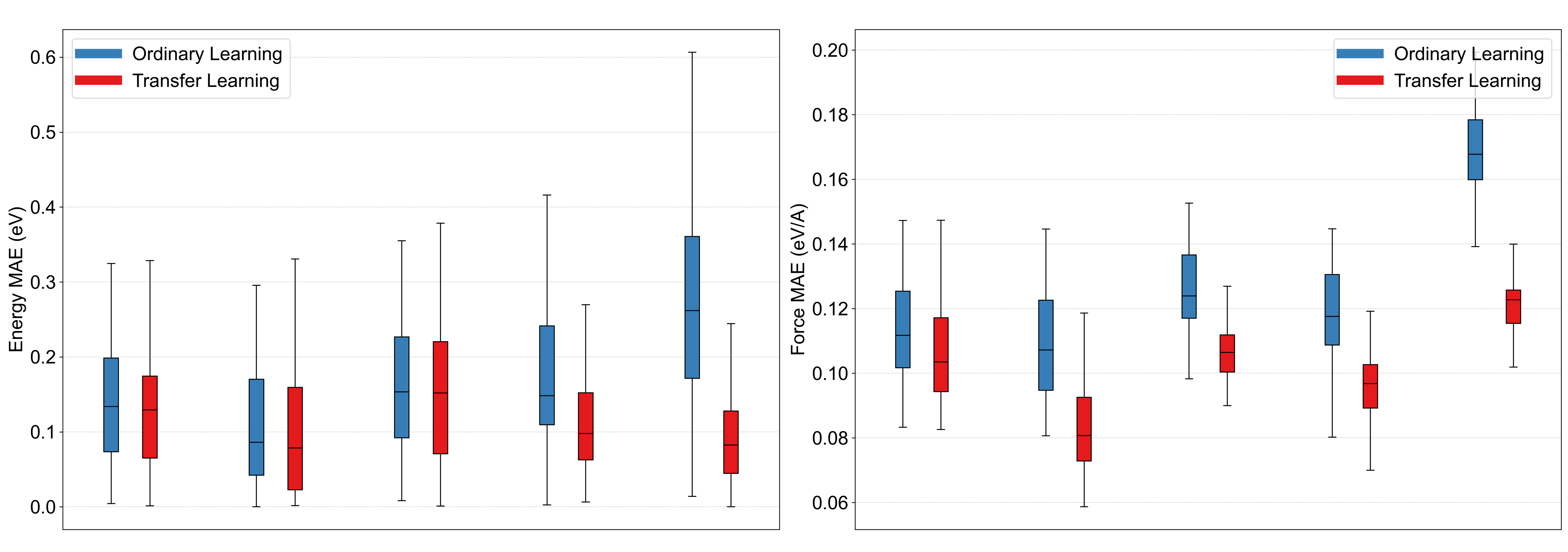}
    \caption{Boxplots representing the distribution of a) energy and b) force errors for five systems outside the training for the first excited state. Two models are compared: transfer learning (red) and training from scratch (blue).}
    \label{fig:state2}
\end{figure*}

\begin{figure*}[htbp] % Use figure* for a wide figure
    \centering
    \includegraphics[width=\textwidth]{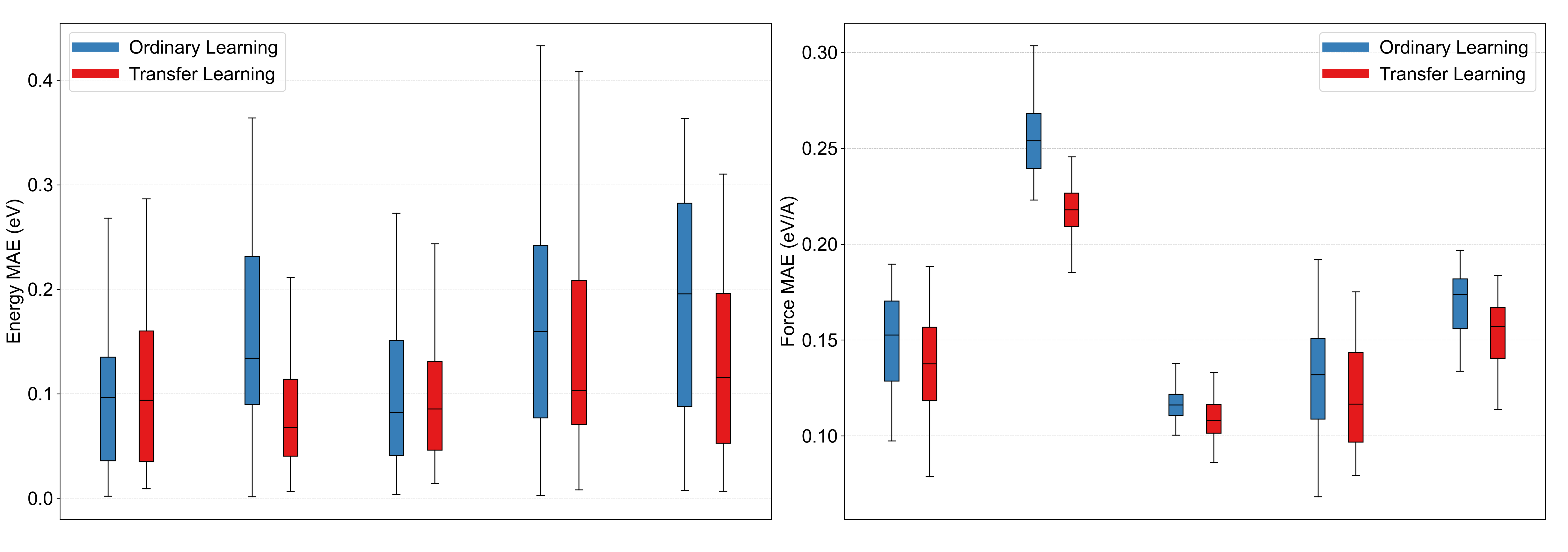}
    \caption{Boxplots representing the distribution of a) energy and b) force errors for five systems outside the training for the second excited state. Two models are compared: transfer learning (red) and training from scratch (blue).}
    \label{fig:state3}
\end{figure*}

\begin{figure*}[htbp] % Use figure* for a wide figure
    \centering
    \includegraphics[width=\textwidth]{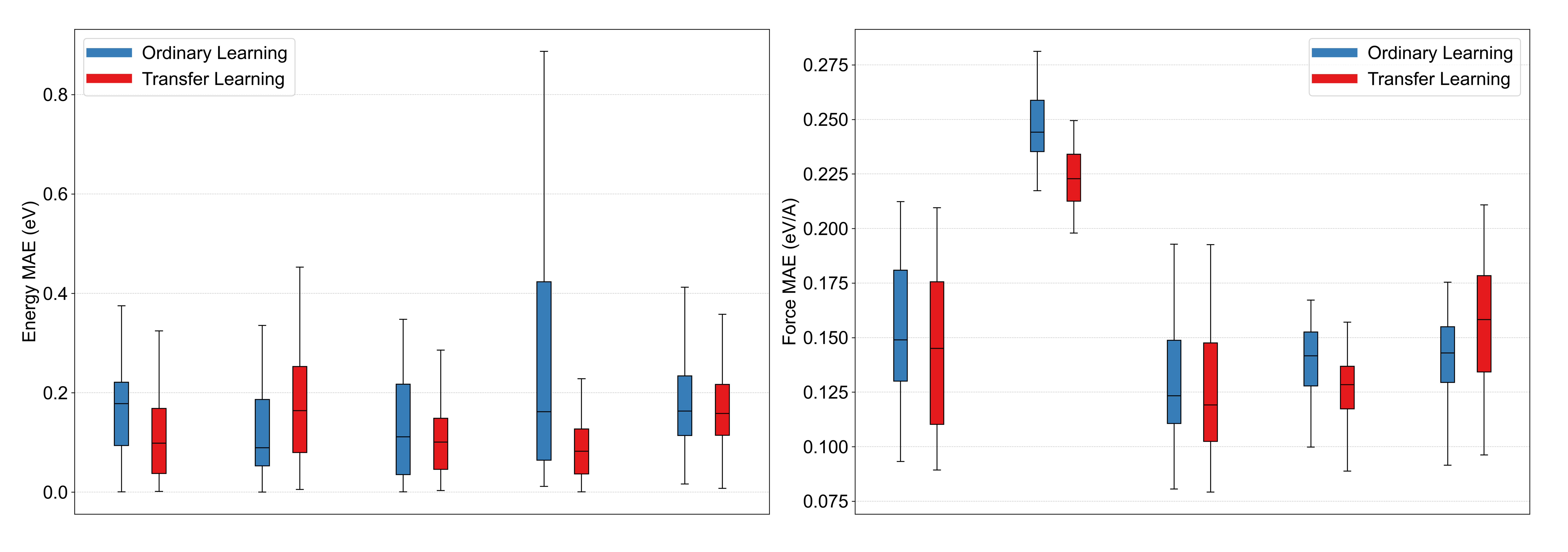}
    \caption{Boxplots representing distribution of a) energy and b) force errors for five systems outside the training for the third excited state. Two models are compared: transfer learning (red) and training from scratch (blue).}
    \label{fig:state4}
\end{figure*}

\begin{figure*}[htbp] % Use figure* for a wide figure
    \centering
    \includegraphics[width=\textwidth]{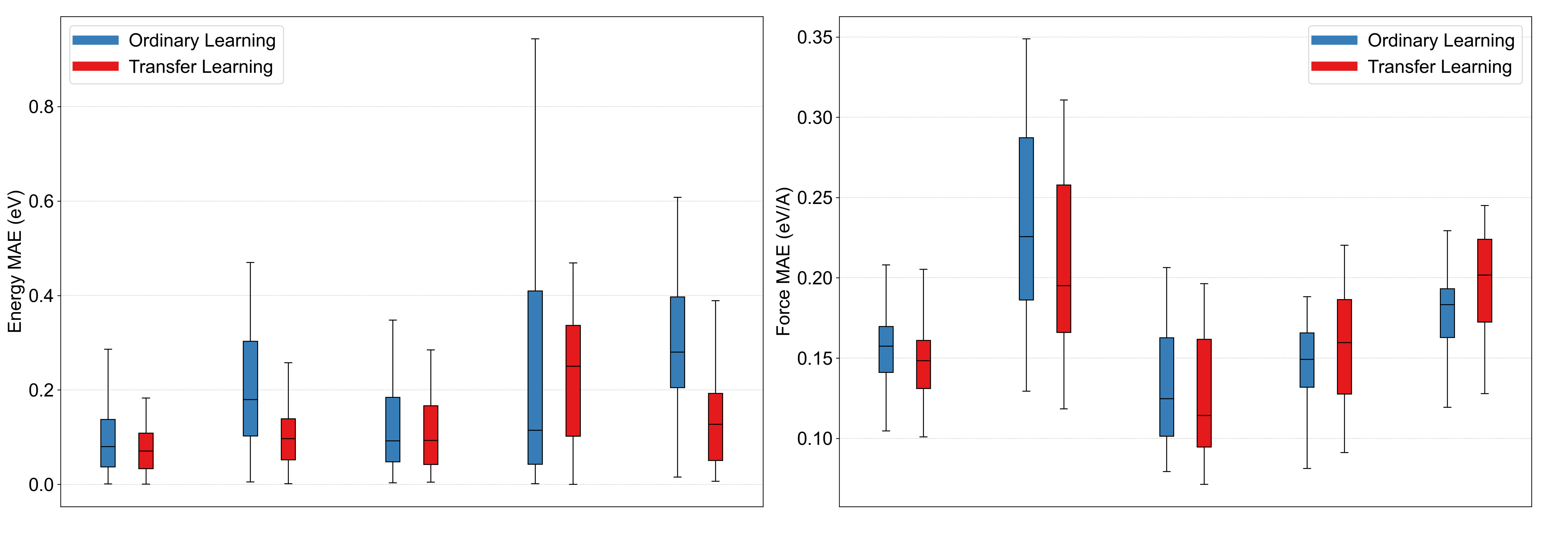}
    \caption{Boxplots representing distribution of a) energy and b) force errors for five systems outside the training for the fourth excited state. Two models are compared: transfer learning (red) and training from scratch (blue).}
    \label{fig:state5}
\end{figure*}

\begin{figure*}[htbp] % Use figure* for a wide figure
    \centering
    \includegraphics[width=\textwidth]{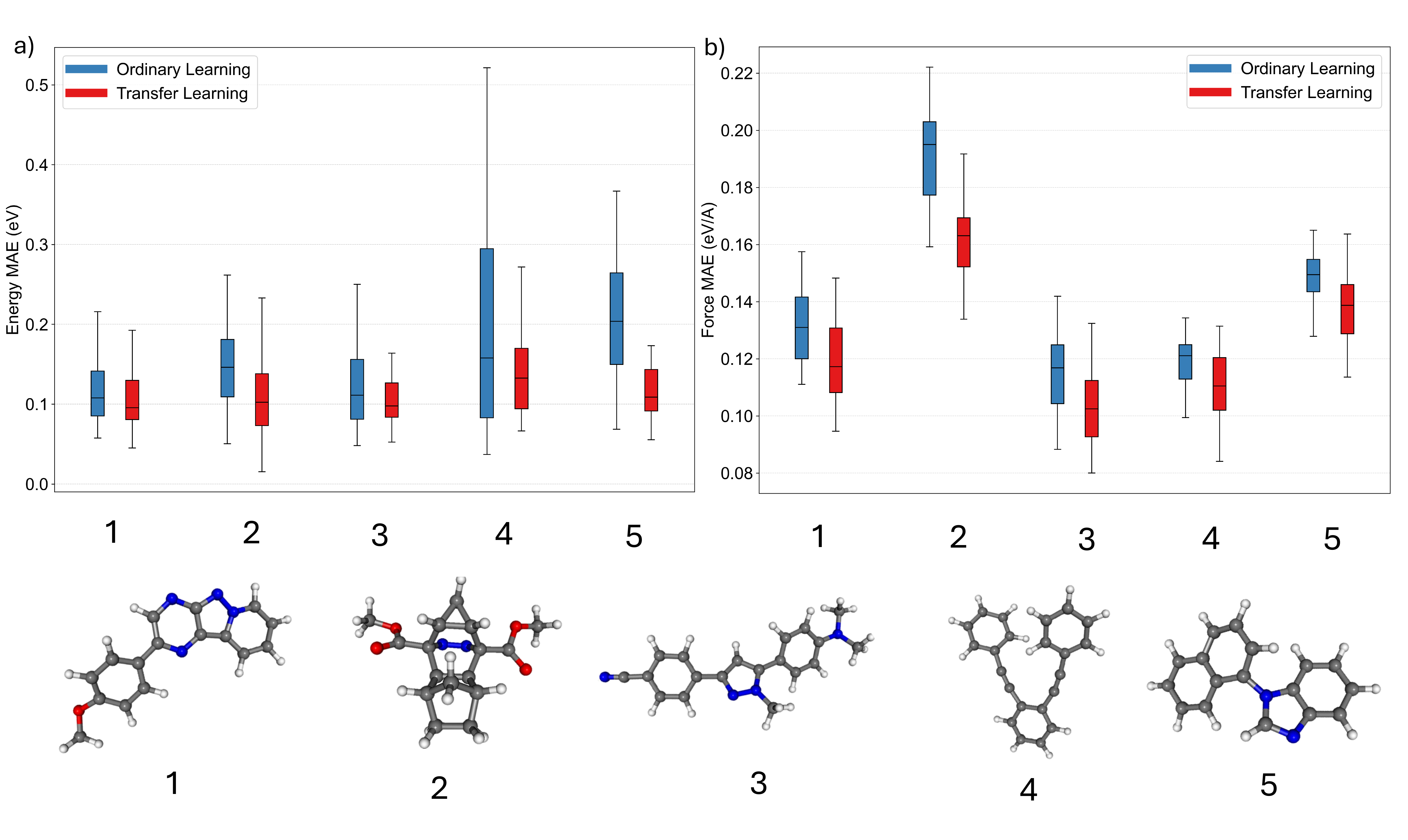}
    \caption{Boxplots representing the distribution of a) energy and b) force errors for five systems outside the training averaging over all states. Two models are compared: transfer learning (red) and training from scratch (blue).}
    \label{fig:avg_state}
\end{figure*}
\FloatBarrier

\section{Phase Free NACs training}
The phase-free loss function is a modification of the mean squared error (MSE) loss designed to handle the arbitrary phase present in nonadiabatic couplings (NACs). In quantum chemistry calculations, NACs are defined only up to an arbitrary phase factor due to the freedom in choosing the wavefunction's global phase. This phase ambiguity complicates direct supervised learning using machine learning models.\cite{akimov2018JPCL,westermayr2019machine} To resolve this, a phase-free loss function originally developed in ref. \citenum{westermayr2020combining} ensures that the learned NACs remain invariant under phase changes. The conventional MSE loss for training the NACs is given by:
\begin{equation}
L_2 = \sum_{i} ||C_{QC, i}^{NAC} - C_{ML, i}^{NAC}||^2,
\end{equation}

which is adapted to account for phase ambiguity in NACs, hence introducing the phase-free loss function. Instead of comparing a single squared error, it evaluates multiple squared errors across different possible sign conventions of the NACs. The modified loss function is defined as:
\begin{equation}
L_{NAC} = \sum_{i,j} \min \left( || C_{QC, ij} - C_{ML, ij} ||^2 ,|| C_{QC, ij} + C_{ML, ij} ||^2 \right)
\end{equation}
The final phase-free loss function then combines the energy and force loss terms with the optimized NAC loss:
\begin{equation}
L_{ph} = a_E \sum_{i} ||E_{QC, i} - E_{ML, i}||^2 + a_F \sum_{i} ||F_{QC, i} - F_{ML, i}||^2 + a_{NAC} L_{NAC}.
\end{equation}
Each term is weighted by a factor $a$. This formulation effectively eliminates the phase dependency of NACs, enabling robust machine learning predictions of excited-state properties. By optimizing over all possible sign conventions, the model ensures that the learned couplings are physically meaningful and independent of arbitrary wavefunction phase choices. The phase-free loss approach also significantly reduces computational costs, as it avoids the need for explicit phase tracking or correction during dataset generation.

\section{Error Breakdown}
The tables below show the error values for the four molecules considered in the Table 1 in the main text broken down by the representation size used. 

\begin{table*}[ht]
\centering
\begin{tabular}{ccccc}
\toprule
\multirow{2}{*}{\textbf{Model}} & \multirow{2}{*}{\makecell{\textbf{Representation}\\\textbf{size}}} & \multicolumn{3}{c}{\textbf{Mean Absolute Error (MAE)}} \\
\cmidrule(lr){3-5}
 &  & \textbf{Energy (eV)} & \textbf{Force (eV/\AA)} & \textbf{Smooth NACS (eV/\AA)} \\
\midrule
\multirow{4}{*}{SchNarc} 
& 16      & 0.0718 & 0.222 & 0.207 \\
& 32            & 0.0525 & 0.134 & 0.147 \\
& 64            & 0.0225 & 0.090 & 0.185 \\
& 128    & 0.0272 & 0.0930 & 0.229 \\
\midrule
\multirow{4}{*}{SPaiNN} 
& 16      & 0.0647 & 0.123 & 0.223 \\
& 32            & 0.0263 & 0.044 & 0.203 \\
& 64            & 0.0072 & 0.0282 & 0.106 \\
& 128    & 0.013 & 0.0532 & 0.117 \\
\midrule
\multirow{4}{*}{E-MACE} 
& 16      & 0.0055 & 0.0375 & 0.103 \\
& 32            & 0.0038 & 0.032 & 0.104 \\
& 64            & 0.0038 & 0.028 & 0.103 \\
& 128     & 0.0032 & 0.0295 & 0.0970 \\
\midrule
\multirow{4}{*}{X-MACE} 
& 16      & 0.0045 & 0.0365 & 0.112 \\
& 32            & 0.0040 & 0.0275 & 0.105 \\
& 64            & 0.0036 & 0.0250 & 0.105 \\
& 128  & 0.0020 & 0.020 & 0.0976 \\
\bottomrule
\end{tabular}
\caption{Mean Absolute Error (MAE) for energies,  forces, and nonadiabatic couplings (NACS) of different models trained on the first three excited states of Ethene}
\label{tab:Ethene}
\end{table*}

\begin{table*}[ht]
\centering
\begin{tabular}{ccccc}
\toprule
\multirow{2}{*}{\textbf{Model}} & \multirow{2}{*}{\makecell{\textbf{Representation}\\\textbf{size}}} & \multicolumn{3}{c}{\textbf{Mean Absolute Error (MAE)}} \\
\cmidrule(lr){3-5}
 &  & \textbf{Energy (eV)} & \textbf{Force (eV/\AA)} & \textbf{Smooth NACS (eV/\AA)} \\
\midrule
\multirow{4}{*}{SchNarc} 
& 16      & 0.159 & 0.456 & 0.493 \\
& 32            & 0.112 & 0.377 & 0.405 \\
& 64            & - & - & - \\
& 128    & 0.0869 & 0.308 & 0.395 \\
\midrule
\multirow{4}{*}{SPaiNN} 
& 16      & 0.153 & 0.450 & 0.441 \\
& 32            & 0.129 & 0.380 & 0.464 \\
& 64            & 0.0732 & 0.282 & 0.395 \\
& 128    & 0.0549 & 0.217 & 0.392 \\
\midrule
\multirow{4}{*}{E-MACE} 
& 16      & 0.143 & 0.280 & 0.266 \\
& 32            & 0.132 & 0.246 & 0.249 \\
& 64            & 0.112 & 0.225 & 0.239 \\
& 128     & 0.0997 & 0.233 & 0.230 \\
\midrule
\multirow{4}{*}{X-MACE} 
& 16      & 0.0979 & 0.234 & 0.271 \\
& 32            & 0.0930 & 0.242 & 0.253 \\
& 64            & 0.0860 & 0.202 & 0.250 \\
& 128  & 0.0863 & 0.213 & 0.236 \\
\bottomrule
\end{tabular}
\caption{Mean Absolute Error (MAE) for energies,  forces, and nonadiabatic couplings (NACS) of different models trained on the first three excited states of the methylenimmonium cation}
\label{tab:methyl}
\end{table*}

\begin{table*}[ht]
\centering
\begin{tabular}{ccccc}
\toprule
\multirow{2}{*}{\textbf{Model}} & \multirow{2}{*}{\makecell{\textbf{Representation}\\\textbf{size}}} & \multicolumn{3}{c}{\textbf{Mean Absolute Error (MAE)}} \\
\cmidrule(lr){3-5}
 &  & \textbf{Energy (eV)} & \textbf{Force (eV/\AA)} & \textbf{Smooth NACS (eV/\AA)} \\
\midrule
\multirow{4}{*}{SchNarc} 
& 16      & 1.786 & 0.399 & - \\
& 32            & 1.644 & 0.408 & - \\
& 64            & 1.506 & 0.433 & - \\
& 128    & 1.124 & 0.407 & - \\
\midrule
\multirow{4}{*}{SPaiNN} 
& 16      & 1.705 & 0.396 & - \\
& 32            & 1.120 & 0.469 & - \\
& 64            & 0.955 & 0.461 & - \\
& 128    & 0.451 & 0.349 & - \\
\midrule
\multirow{4}{*}{E-MACE} 
& 16      & 0.149 & 0.153 & - \\
& 32            & 0.123 & 0.134 & - \\
& 64            & 0.103 & 0.118 & - \\
& 128     & 0.0973 & 0.107 & - \\
\midrule
\multirow{4}{*}{X-MACE} 
& 16      & 0.109 & 0.135 & - \\
& 32            & 0.0986 & 0.122 & - \\
& 64            & 0.0930 & 0.114 & - \\
& 128  & 0.0904 & 0.108 & - \\
\bottomrule
\end{tabular}
\caption{Mean Absolute Error (MAE) for energies,  forces, and nonadiabatic couplings (NACS) of different models trained on the first five excited states of Chromphores}
\label{tab:chromophores}
\end{table*}

\begin{table*}[ht]
\centering
\begin{tabular}{ccccc}
\toprule
\multirow{2}{*}{\textbf{Model}} & \multirow{2}{*}{\makecell{\textbf{Representation}\\\textbf{size}}} & \multicolumn{3}{c}{\textbf{Mean Absolute Error (MAE)}} \\
\cmidrule(lr){3-5}
 &  & \textbf{Energy (eV)} & \textbf{Force (eV/\AA)} & \textbf{Smooth NACS (eV/\AA)} \\
\midrule
\multirow{4}{*}{SchNarc} 
& 16      & 0.0998 & 0.272 & 0.128 \\
& 32            & 0.0755 & 0.201 & 0.192 \\
& 64            & 0.0484 & 0.128 & 0.124 \\
& 128    & 0.0509 & 0.082 & 0.136 \\
\midrule
\multirow{4}{*}{SPaiNN} 
& 16      & 0.0534 & 0.144 & 0.138 \\
& 32            & 0.0394 & 0.0996 & 0.124 \\
& 64            & 0.0276 & 0.0620 & 0.132 \\
& 128    & 0.0156 & 0.0463 & 0.106 \\
\midrule
\multirow{4}{*}{E-MACE} 
& 16      & 0.0160 & 0.0630 & 0.0950 \\
& 32            & 0.0113 & 0.0472 & 0.0891 \\
& 64            & 0.00787 & 0.0360 & 0.0854 \\
& 128     & 0.00687 & 0.0332 & 0.0868 \\
\midrule
\multirow{4}{*}{X-MACE} 
& 16      & 0.0172 & 0.0671 & 0.100 \\
& 32            & 0.0129 & 0.0515 & 0.0964 \\
& 64            & 0.00884 & 0.0388 & 0.0882 \\
& 128  & 0.00664 & 0.0315 & 0.0869 \\
\bottomrule
\end{tabular}
\caption{Mean Absolute Error (MAE) for energies,  forces, and nonadiabatic couplings (NACS) of different models trained on the first five excited states of Butene}
\label{tab:methyl}
\end{table*}

\providecommand*{\mcitethebibliography}{\thebibliography}
\csname @ifundefined\endcsname{endmcitethebibliography}
{\let\endmcitethebibliography\endthebibliography}{}